\documentstyle[prd,aps,preprint]{revtex}

\let\ssection=\section
\renewcommand{\section}{\setcounter{equation}{0}\ssection}
\def\sqr#1#2{{\vcenter{\hrule height.#2pt\hbox{\vrule width.#2pt
-1zheight#1pt \kern#1pt \vrule width.#2pt}\hrule height.#2pt}}}

\newcommand{\be}{\begin{equation}}
\newcommand{\ee}{\end{equation}}
\newcommand{\ben}{\begin{eqnarray}}
\newcommand{\een}{\end{eqnarray}}
\newcommand{\bec}{\begin{center}}
\newcommand{\eec}{\end{center}}

\begin{document}

\draft
\widetext

\title{Reconstructing the inflaton potential for an almost flat COBE
spectrum}

\author{Eckehard W.~Mielke\footnote{Electronic address:
pke27@rz.uni-kiel.d400.de [1.9.-19.12.95: c/o A.~Mac{\' i}as,
visit@xanum.uam.mx]} and Franz E.~Schunck\footnote{Electronic address:
fs@thp.uni-koeln.de}}

\address{Institute for Theoretical Physics, University of Cologne,
D-50923 K\"oln, Germany}

\date{\today}

\maketitle

\begin{abstract}
Using the Hubble parameter as new `inverse time'
coordinate ($H$-formalism), a new method of reconstructing the inflaton
potential is developed also using older results which, in 
principle, is applicable to any order of the slow-roll approximation.
In first and second order, we need three
observational data as inputs: the scalar spectral
index $n_s$ and the amplitudes of the scalar and the tensor spectrum.
We find constraints between the values of $n_s$ and the corresponding values
for the wavelength $\lambda $. By imposing a dependence
$\lambda (n_s)$, we were able to reconstruct and visualize
inflationary potentials which are compatible with recent COBE and
other astrophysical observations.
From the reconstructed potentials, it becomes clear that one cannot
find only one special value of the scalar spectral index $n_s$.
\end{abstract}
\bigskip\bigskip

\narrowtext

\section {\bf Introduction}

The observations by the Cosmic Background Explorer (COBE)
may have shown some hints on the nature of the
inflation driven by the vacuum energy \cite{smoot,smoot2}.
In theoretical models this vacuum energy is
simulated by the self-interaction potential of a scalar inflaton
field. The amplitudes of the scalar and
the tensor perturbations and the scalar and the gravitational spectral
index of the background radiation are astrophysically observable.
Inflationary models \cite{lidlyt,barlid,lidlyt2} show that the
{\em scalar spectral index} $n_s$ takes values both between zero and one
and, for some models, beyond 1.
Recent data \cite{silk} provide now a value of $n_s$ between 1.1 and
1.59 which is slightly beyond the Harrison-Zel'dovich spectrum;
cf.~recently values in \cite{bennett}.

The previous reconstructions of inflationary 
potentials \cite{hodg,cope,cope2} have used both
approximations and exact potentials depending on the wavelength
$\lambda $. The value of the
potential at a special wavelength $\lambda_0$ or at a special value of the
scalar field together with its first and second derivatives could be
reconstructed. Hence,
experimental data at different wavelengths determine, in this way, the
form of the inflationary potential.

In \cite{schunck} we have found the general {\em exact} inflationary solution
depending on the Hubble constant $H$, the `inverse time', and were able
to classify a regime in which inflationary potentials are viable.
In this paper, we apply this $H$-formalism to the first and second order
perturbation formalism and reconstructed a phenomenologically viable
inflationary
potential. The construction of the graceful exit function $g(H)$ is
the essential point of our new method. Our function $g(H)$,
parametrized by $n_s$,
determines the inflaton potential and the exact Friedman type 
solution. From this differential equation we
are able to present three different type of potentials for $n_s=1$,
$n_s>1$, and $0<n_s<1$. Moreover,
for each regime of $n_s$, we find a different dependence on the
wavelength $\lambda (n_s)$. This can be important for future
observations.

\section {\bf General metric of a spatially flat inflationary universe}

For a rather general class of inflationary models the Lagrangian density reads
\be
{\cal L} = \frac{1}{2 \kappa } \sqrt{\mid g \mid}
 \Biggl ( R
   + \kappa \Bigl [ g^{\mu \nu } (\partial_\mu \phi ) (\partial_\nu \phi )
   - 2 U(\phi ) \Bigr ] \Biggr )  \; , \label{lad}
\ee
where $\phi $ is the scalar field and $U(\phi )$
{\em the self-interaction potential}. We use natural units with $c=\hbar =1$.
A constant potential $U_0= \Lambda / \kappa $ would simulate the
cosmological constant $\Lambda $.

For the {\em flat} ($k=0$) Robertson-Walker metric  
\be
ds^2 = dt^2 - a^2(t) \left [ dr^2 + r^2 \left (
       d\theta^2 + \sin^2 \theta d \varphi^2 \right ) \right ]
\; , \nonumber \\
\ee
the evolution of the generic inflationary model (\ref{lad}) is
determined by the autonomous first order equations 
\ben
\dot H & = & \kappa U(\phi ) - 3H^2 \; , \label{doth} \\
\dot \phi & = & \pm \sqrt {\frac {2}{\kappa }}
  \sqrt{3H^2 - \kappa U(\phi )} \; . \label{dotphi}
\een
This system corresponds to the Hamilton-Jacobi Eqs.~(2.1) and (2.2)
of Ref.~\cite{cope2}.
However, by introducing the Hubble expansion 
rate $H:=\dot  a(t)/a(t)$ as the {\em new} `inverse time'
{\em coordinate} we \cite{schunck} found the general solution:
\ben
t & = & t(H) = \int
   \frac {dH}{\kappa \widetilde U - 3 H^2}  \label{tH} \; , \\
a & = & a(H) = a_0 \exp \left (
   \int \frac {H dH}{\kappa \widetilde U - 3 H^2} \right )  \label{aH}
\; . \\
ds^2 & = & 
\frac {dH^2}{\left (\kappa \widetilde U - 3 H^2 \right )^2}
 - a_0{}^2 \exp \left ( 2
   \int \frac {H dH}{\kappa \widetilde U - 3 H^2} \right ) \times  \nonumber \\
 & &   \left [ d r^2 + r^2 \left (
    d \theta^2 + \sin^2 \theta d \varphi^2 \right ) \right ] \; , \\
\phi & = & \phi (H) = \mp \sqrt {\frac{2}{\kappa }}
 \int \frac {dH}{\sqrt{3H^2-\kappa \widetilde U}} \; , \label{phiH}
\een
where $\widetilde U = \widetilde U (H) := U(\phi (t(H)))$ is the
{\em reparametrized} inflationary potential.

Since the singular case $\widetilde U = 3H^2/\kappa $ leads to the
{\em de Sitter inflation}, we use in explicit models the ansatz
\be
\widetilde U(H) = \frac {3}{\kappa } H^2 + \frac {g(H)}{\kappa} \label{uh}
\ee
for the potential, where $g(H)$ is a nonzero function for the
{\em graceful exit}. Our $H$-formalism will facilitate considerably the 
reconstruction of the inflaton potential to any order, as we will 
demonstrate in the following.

A classification of all allowed inflationary
potentials and scenarios has recently been achieved by Kusmartsev et
al.~\cite{kusm}
via the application of {\em catastrophy theory} to the Hamilton-Jacobi type
equations (\ref{doth}) and (\ref{dotphi}).

\section {\bf Density perturbations}

For a long time
one thought that the spectrum of {\em density perturbations} is described by
the scale-invariant Harrison-Zel'dovich form
\cite{guthpi,hawking,starobinsky}. But new observations by COBE
\cite{smoot} show the possibility of small deviations.
In Ref.~\cite{cope,cope2} the slow-roll approximation is specified by the
three parameters $\epsilon$, $\eta $, and $\xi $
defined as follows:
\ben
\epsilon & := &
  \frac {2}{\kappa } \left ( \frac {H'}{H} \right )^2
 = - {{g}\over{H^2}}  \label{eps} \; , \\
\eta & := &
  \frac {2}{\kappa } \frac {H''}{H}
 = 3 -{\kappa\over 2 H}{{d\widetilde U}\over{dH}}
 = -{dg\over{dH^2}} \; , \\
\xi & := & {2\over\kappa} {{H'''}\over{H'}} =
\eta - \frac {2\eta '}{\sqrt {2 \kappa \epsilon }}
= -{dg\over{dH^2}} - 2 H^2\,{{d^2g}\over{(dH^2)^2}} \label{xi} \; .
\een
From our $H$-formalism \cite{schunck}, the relations for the graceful
exit function $g(H)$ on the right hand side of (\ref{eps})-(\ref{xi})
follow quite generally. These slow-roll parameters 
effectively provide a Taylor expansion of the graceful exit function.
The minus sign in
$\sqrt {\epsilon } = - \sqrt{2/\kappa } H'/H$ is necessary in order to be
consistent with the choice $\dot \phi >0$, cf.~\cite{cope2}.

In general, these parameters are scale-dependent and have to be
evaluated at the horizon. The parameter
$\epsilon $ describes the relation between the kinetic and the total
energy, whereas $\eta $ is a measure for the relation between the
``acceleration'' of the scalar field and its ``curvature-depending velocity''.
In the slow-roll approximation, all three parameters
are {\em small quantities}. Actually, the phase of acceleration
($\ddot a>0$) is now equivalently to the condition $\epsilon <1$.

The amplitudes of scalar
and transverse-traceless tensor perturbations \cite{lytste,lidlyt}
are given in first order slow-roll approximation \cite{lidlyt} by
\ben
P^{\frac {1}{2}}_{{\cal R}} (\hat k) & = &
 \left ( \frac {H^2}{4\pi \mid H' \mid } \right )
 \Biggl |_{aH=\hat k} \; , \label{scamp} \\
P^{\frac {1}{2}}_{g} (\hat k) & = &
 \left ( \frac {H}{2\pi } \right ) \Biggl |_{aH=\hat k} \label{teamp} \; ,
\een
where ${\cal R}$ denotes the perturbation in the spatial curvature,
$H'=dH/d\phi $, and $\hat k:=2\pi/\lambda $ the wave number. The
expressions on the
right hand side have to be evaluated at that comoving scale $\hat k$
which is leaving the horizon during the inflationary phase.
The {\em scalar} and the {\em gravitational spectral indices} in first
order approximation read
\ben
n_s & := & 1 + \frac {d \ln P_{{\cal R}}}{d \ln\hat k}
= 1 - 4 \epsilon + 2 \eta \; , \label{ns} \\
n_g & := & \frac {d \ln P_{g}}{d \ln\hat k}
= - 2 \epsilon \le 0 \; . \label{ng}
\een
The last condition is fulfilled, because $\epsilon >0$ or $g<0$, respectively.

The relation of the {\em wavelength} $\lambda $ to the scalar field
\cite{cope,cope2}
\be
\frac {d\lambda }{d\phi } = \lambda \frac{H}{H'}
  \frac {\kappa }{2} (1-\epsilon )
\ee
converts, in the $H$-formalism, exactly to
\be
\frac {d\ln \lambda }{dH^2} =
  \frac{1}{2 H^2} \left (\frac {1}{\epsilon } - 1 \right ) =
 - \frac{1}{2} \left (\frac {1}{H^2} + \frac {1}{g} \right )
\; . \label{lamHgen}
\ee
By using the slow-roll condition $\epsilon <<1$, the $H$-dependence
simplifies to \cite{hodg}
\be
\frac {d\lambda }{d H} \simeq \frac {\lambda }{H \epsilon } \label{lamH} \; .
\ee

\section {\bf Reconstructing the inflaton potential in first order}

From Eq.~(\ref{ns}) we find the differential equation for the
graceful exit function
\be
\frac {dg}{dH} = (1-n_s) H + \frac {4g}{H} \; ,
\ee
which has the solution
\be
g(H) = \Delta \, H^2 - A H^4  \label{gh} \; ,
\ee
where we abbreviated the deviation from the flat spectrum, i.e., $n_s=1$,
by $\Delta:= (n_s-1)/2$.
For the integration constant $A$ we find the $n_s$-dependent reality 
condition
\be
0 \le \frac {\Delta }{H^2} < A  \; ,
\ee
provided $n_s \ge 1$. In second order perturbation, the condition
\be
\xi = \Delta - 2 A H^2 << 1
\ee
would arise.

We can distinguish an
{\em inflationary} and a {\em Friedmann era} of spacetime for the
potential
\be
\widetilde U(H) = \frac {1}{\kappa }
\biggl [ (\Delta +3) H^2 - A H^4 \biggr ] \; . \label{xxx}
\ee
The inflation starts at the `inverse time' $H_1=\sqrt{\Delta /A}$
(which means $\kappa \widetilde U=3H^2$),
it ends at $H_2=\sqrt{(\Delta +1)/A}$ (which corresponds to
$\kappa \widetilde U=2H^2$),
whereas for $H_3=\sqrt{(\Delta +3)/A}$ we have $\widetilde U(H)=0=U(\phi )$.
Of course, only the inflationary and the beginning of the Friedmann
parts are physically relevant. After
this, the approximation is no longer valid.
For $0 \le H \le H_{1}$, the dilaton field would become a ghost (or one
could construct gravitational instantons, following \cite{muk}).
The duration of inflation determines a range for $A$
\be
\Delta < A H^2 < \Delta +1  \; . \label{freea}
\ee

If $\Delta <0$, i.e., $n_s<1$, and $A>0$, the inflationary phase
exists but does no longer start at the point
$\kappa \widetilde U=3H^2$. If $\Delta +1$ is negative (i.e.,
$n_s \le -1$), the constant $A$ has to become negative, too, in order to
allow an inflationary phase.  Hence, in
all such cases, we find an inflationary phase.

Via $g =: -d\widetilde{W}/dH$, the solution (\ref{gh})
corresponds to the non-Morse function
\be
\widetilde{W}_{\rm deform} = {A\over 5}\, H^5 - \frac {\Delta }{3} H^3 + B
\; ,
\ee
($B$ is an integration constant) and therefore belongs to the 4th
Arnold class $A_4$, see our recent bifurcation analysis of inflation in
Ref.~\cite{kusm}.

\subsection {\bf The Harrison-Zel'dovich potential}

The {\em flat spectrum} of
Harrison-Zel'dovich is obtained for $n_s=1$.
From (\ref{gh}), (\ref{uh}), and (\ref{tH}) we get the Hubble expansion rate
\be
H = \Bigl [ 3 A (t+C_1) \Bigr ]^{-1/3} \; ,
\ee
whereas the scale factor reads
\be
a(t) = a_0 \exp \left [ (3A)^{-1/3} \frac {3}{2}
(t+C_1)^{2/3} \right ] \; .
\ee
The scalar field is then given by
\be
\phi (t) + C_3 =
\pm \sqrt { \frac {2}{A \kappa } }
 \Bigl [ 3 A (t+C_1) \Bigr ]^{1/3}
= \pm \sqrt { \frac {2}{A \kappa } } \frac {1}{H} \; .
\ee
The corresponding potential\footnote{Eq.~(7.14) in \cite{schunck} is
misprinted; the correct potential reads:
$$
U(\phi ) = \frac{1}{\kappa } \left [ \frac {A\kappa }{8} (2-n)^2 
  (\phi + C_3)^2 \right ]^{2/(2-n)}
  \left (3 - \frac {8}{\kappa (2-n)^2 (\phi + C_3)^2} \right ) \; .
$$}
\be
U(\phi ) =
{6 \over {A\kappa^2} }\, (\phi +C_3)^{-4} \, \left[ (\phi +C_3)^2 
 - {2 \over {3\kappa} } \right ]  \label{un}
\ee
describes the flat Harrison-Zel'dovich spectrum.

The more general ansatz
\be
g(H) = - A \, H^{n} \; ,
\ee
where $n$ is real and $A$ a positive constant of dimension 
$length^{n-2}$, leads to several known and new solutions, 
cf.~\cite{schunck}.

\subsection{\bf Inflationary potential with an almost flat spectrum: $n_s>1$}

For $n_s>1$, i.e., $\Delta :=(n_s - 1)/2>0$, and $A>0$, we
find from (\ref{gh}) and (\ref{tH})-(\ref{phiH}) the solution
\ben
t & = & - \frac {1}{\Delta H} + \sqrt {\frac{A}{\Delta ^3}} \;
   \mbox{arcoth} \left [ \sqrt {\frac {A}{\Delta }} H \right ]
\; , \label{tns} \\
a(H) & = & a_0
   \left ( \frac {A H^2}{A H^2-\Delta } \right )^{1/(2\Delta )}
\; , \label{ans} \\
\phi(H) & = & \mp\sqrt{{2\over{\kappa \Delta}}}
\arcsin\left(-\sqrt{{\Delta \over A}}\, {1\over H}\right) \; , \label{pp}\\
H(\phi ) & = & - \sqrt {\frac {\Delta }{A}} \frac {1}{
     \sin \left (\mp \sqrt {\frac {\Delta  \kappa }{2}} \phi \right )} \; .
\een
The potential $U(\phi )$ reads
\be
U(\phi ) = \frac {\Delta }{\kappa A}
    \frac {1}{\sin^2 \left (\mp \sqrt {\frac {\kappa \Delta }{2}}
    \phi \right )}
    \left ( \Delta +3
  - \frac {\Delta }{
     \sin^2 \left (\mp \sqrt {\frac {\kappa \Delta }{2}}
    \phi \right )} \right )
\; .  \label{un1}
\ee
We recognize that the limit $\Delta \rightarrow 0$ is not singular for
$a,\phi ,H,U$; for $t(H)$ the limit is indeterminate (confirmed by
MATHEMATICA and
MAPLE), because of the restricted definition range of arcoth.

We investigate now the case where $n_s=const$.
The inflaton starts in an extremum of the potential $U(\phi )$
($\hat = \kappa \widetilde U=3H^2$).
Then, depending on $n_s$ we find two types of behavior of the potential. The
local {\em extrema} of the potential occur at
\ben
\phi_{1} & = & \sqrt {\frac {2}{\kappa \Delta }} (m + 1)
  \frac {\pi }{2} \; ,  \qquad  m = 0, 1, 2, \cdots\\
\phi_{2} & = & \sqrt {\frac {2}{\kappa \Delta }}
  \arcsin \left ( \sqrt {\frac {2\Delta }{3+\Delta }} \right ) \; .
\een
For $\Delta  < 3$ or $n_s < 7$, respectively, $\phi_1$ is always a
minimum at the beginning of inflation followed by a maximum at $\phi_2$.
But for $n_s \ge 7$, the maximum at $\phi_2$ disappears and $\phi_1$,
the previous minimum, becomes a maximum.
The potentials for $n_s < 7$ belongs to the ``old'' inflationary
theory, whereas for $n_s \ge 7$ we would find the ``new'' inflationary
potentials.
Fig.~\ref{fig.0} shows the possible inflationary potentials for
$n_s=1.01$ ... 2.1. The inflationary potential which will be measured
is then a ``way on this rug''. Fig.~\ref{fig.1} and
Fig.~\ref{fig.2} show two cross-sections within the ``rug''.

\subsection{\bf Inflationary potential with an almost flat spectrum: $n_s<1$}

For $0<n_s<1$, we have to distinguish two cases:
a) $\Delta <0$ and $A<0$, b) $\Delta <0$ and $A>0$. In both cases, the
reality condition $g<0$ can be satisfied. The solution of case a) is
given by Eqs.~(\ref{tns})-(\ref{un1}). For case b), we find
\ben
t & = & - \frac {1}{\Delta H} + \sqrt {- \frac{A}{\Delta ^3}} \;
   \arctan \left [ \sqrt {- \frac {A}{\Delta }} H \right ]
\; ,  \\
\phi(H) & = & \mp\sqrt{-{2\over{\kappa \Delta}}}\;
\ln \left(\frac {1}{H} \left [ \sqrt{-{\Delta \over A}} +
  \sqrt {- \frac{\Delta }{A} + H^2} \right ] \right ) \nonumber \\
 & = & \mp\sqrt{-{2\over{\kappa \Delta}}}\;
\mbox{arsinh} \left(-\sqrt{-{\Delta \over A}}\, {1\over H} \right)
\; , \label{pp2} \\
H(\phi ) & = & \sqrt {-\frac {\Delta }{A}}
 \frac {\exp \left (\mp \sqrt {\frac {\kappa \Delta }{2}} \phi \right )}
 {\exp \left (\mp \sqrt {- 2 \kappa \Delta } \phi \right ) - 1}
 \; , \\
U(\phi ) & = & - \frac {4 \Delta }{A \kappa }
 \frac {\exp \left (\mp \sqrt {- 2 \kappa \Delta }
        \phi \right )}
     {\left ( \exp \left (\mp \sqrt {- 2 \kappa \Delta }
      \phi \right ) - 1 \right )^2} 
     \left ( \Delta + 3
  + \frac {4 \Delta \exp \left (\mp \sqrt {- 2 \kappa \Delta }
        \phi \right )}
     {\left ( \exp \left (\mp \sqrt {- 2 \kappa \Delta }
      \phi \right ) - 1 \right )^2}
 \right ) \; , \label{un2}
\een
and $a(H)$ is given by (\ref{ans}). Related inflationary 
solutions parametrized by $n_s$ below and above one have been obtained 
with a different method in Ref.~\cite{carr}.

\section{\bf Inflationary potential parametrized by the wavelength}

Because the scalar field is not observable, we also specify the
dependence of the potential on the wavelength $\lambda$.
From (\ref{lamH}) together with (\ref{eps}) and (\ref{gh}), we get
\be
\frac {\lambda }{\lambda_0} \simeq \left ( 1 - \frac {\Delta }{A H^2}
   \right )^{1/(2\Delta )}  \label{wave1} \quad , \qquad
H^2 \simeq \frac {\Delta }{A} \left ( 1 -
  \left [ \frac {\lambda }{\lambda_0} \right ]^{2\Delta }
  \right )^{-1} \; , \label{wave2}
\ee
where $\lambda_0$ is an integration constant.
The value of the inflationary potential (\ref{xxx}) is determined by
\be
U(\lambda , A, n_s) \simeq
\frac {\Delta }{\kappa A (1-[\lambda/\lambda_0]^{2\Delta })}
\left ( \Delta + 3 - \frac {\Delta }{1-[\lambda/\lambda_0]^{2\Delta }}
\right )  \; . \label{un3}
\ee

For $n_s=1$ and $A>0$, we obtain the solution
\ben
H^2 & \simeq & - \frac {1}{2A\ln [\lambda/\lambda_0] } \; , \label{wave3} \\
U   & \simeq & - \frac {1}{2 \kappa A \ln [\lambda/\lambda_0] } \left (
3 + \frac {1}{2\ln [\lambda/\lambda_0] } \right ) \; . \label{un4}
\een
From (\ref{un}), (\ref{un1}), (\ref{un2}), (\ref{un3}), and
(\ref{un4}) we recognize that $A$ merely scales the potential.
From the exact equation (\ref{lamHgen}) we find
$\lambda /\lambda_0 = \exp (-1/(2AH^2))/H$ for $n_s=1$.

From (\ref{wave2}) or (\ref{wave3}), respectively, and the scalar
field solutions (\ref{pp}) and
(\ref{pp2}), we can deduce the following reality conditions:
\ben
0 < n_s < 1 \; , \quad A \neq 0 \quad & \Longleftrightarrow & \quad
[\lambda/\lambda_0]^{2\Delta } > 1  \label{cond1} \\
1 < n_s < \infty \; , \quad A \neq 0   \quad & \Longleftrightarrow & \quad
0 < [\lambda/\lambda_0]^{2\Delta } < 1  \label{cond2} \\
n_s = 1 \; , \quad A>0  \quad & \Longleftrightarrow & \quad
0 < [\lambda/\lambda_0] < 1  \label{cond3}
\een
These conditions determine a possible dependence
$\lambda =\lambda (n_s)$ which may be seen in future
observations.
The result (\ref{cond1})-(\ref{cond3}) for such a functional
dependence is clarified by three examples. Let us suppose, the behavior is
$\lambda/\lambda_0 (n_s) := n_s^2 \; ;$ then the potential $U(\phi )$
has the shape shown in Fig.~\ref{fig.3}. For the functional
dependence $\lambda/\lambda_0 (n_s) := 1/n_s^2 \; ,$
the potential of Fig.~\ref{fig.4} results.
For $\lambda/\lambda_0 (n_s) := 1/\sqrt{n_s}$, we find Fig.~\ref{fig.5}.

\section {\bf Second-order approximation}

Our $H$-formalism allows, in principle, to determine $g(H)$ to arbitrary 
high order in the Taylor expansion. In second order, we obtain a nonlinear
second order equation for the graceful exit function.

The second-order result of the scalar spectral index is
\be
1-n_s = 4 \epsilon - 2 \eta + 8 (1+C) \epsilon^2 -
(6+10 C) \epsilon \eta + 2 C \epsilon \xi  \; , \label{diff}
\ee
where $C=-2+\ln (2) +\gamma \sim -0.73$ and $\gamma \sim 0.577$ is the
Euler constant \cite{stelyt}. In the $H$-formalism, Eq.~(\ref{diff})
converts into the nonlinear second order equation
\be
\Delta = 2 \frac {g}{y} - g'
- 4 (1+C) \left ( \frac {g}{y} \right  )^2
+ (3+4C)\frac {g}{y} g' - 2 C g g''  \label{g2y}
\ee
for the graceful exit function $g$, where $y:=H^2$ and $'=d/d(H^2)$.
In terms of $\epsilon = -g/y$, cf.~(\ref{eps}),
we can rewrite this condition as
\be
2 C \epsilon\, {\buildrel{\bullet\bullet}\over\epsilon} - 
(2C+3)\epsilon\, {\buildrel\bullet\over\epsilon}
- {\buildrel\bullet\over\epsilon}
+ \epsilon^2  + \epsilon + \Delta =0  \; , \label{diff2}
\ee
where $\bullet \hat= d/d\ln y$.
Equation (\ref{g2y}) has the exact $\Delta$--dependent solution 
\be
g = \left ({1\over2} \pm \sqrt{{1\over4} - \Delta}\right) y
  \; . \label{g1}
\ee
The solution with the plus sign can be ruled out because
we require $g<0$, while the solution with the minus sign possesses an
inflationary part if and only if $n_s<1$. The potential for this $g$
was already constructed \cite{schunck}, it belongs to the class of
power-law models:
\be
U(\phi ) = \frac {3-A}{\kappa } C_3{}^2
 \exp (\pm \sqrt{2 \kappa A}\; \phi ) \; .
\ee
Note, however, that in second order the integration constant
$A={1\over2} - \sqrt{{1\over4} - \Delta}$ is
now fixed by the observational data for the scalar spectral index $n_s$.
Because of the nonlinearity of (\ref{g2y}), further solutions exist. 
In order to obtain more inflaton potentials $U(\phi)$ in second order, 
we rewrite (\ref{phiH}) into the form
\be
\frac {d \phi}{dy} =
\mp \sqrt{\frac {2}{\kappa }} \frac {1}{\sqrt{-4 g y}} \; . \label{4gy}
\ee
This equation together with
(\ref{g2y}) combines to a coupled system which can be solved
numerically, for example by using MATHEMATICA. The potential
$U(\phi )$ is given by the parametric solution
$\{ \phi (H), U(H)=(3H^2+g)/\kappa \}$; see Fig.~\ref{fig.6}.

\section{Remarks}

Using the $H$-formalism, in first order, we were able to present three
alternative dependences of the potential: $U(H)$, $U(\phi )$, and
$U(\lambda )$. In order to have no arbitrary integration constant, we need
three observables as input: the two amplitudes for the tensor and the scalar
spectrum and the scalar spectral index $n_s$, each time at the wavelength
$\lambda_0$ under consideration.
From $P^{\frac {1}{2}}_{g}$ at $\lambda_0$, one finds the Hubble
parameter
\be
H^2(\lambda_0) = 4 \pi^2 P_g
\ee
and, from $P^{\frac {1}{2}}_{{\cal R}}$ at $\lambda_0$, the
Harrison-Zel'dovich constant
\be
A(\lambda_0) =
\frac {\sqrt{2}}{4 \pi \sqrt {\kappa}} \frac {1}{P_{{\cal R}}}
+ \frac {\Delta }{4 \pi^2} \frac {1}{P_g}
\ee
follows. From (\ref{pp}) for $n_s>1$ or
from (\ref{pp2}) for $n_s<1$, respectively, the scalar field
$\phi (\lambda_0)$ is obtained and, hence, $U(H(\phi (\lambda_0)))$.
Equation (\ref{wave2}) is the consistency condition for the observed
quantities. Equations (\ref{cond1})-(\ref{cond3}) give, up to first order, some
restrictions for the $\lambda $-$n_s$ relation.

In second order, we were able to present an exact solution for
$n_s<1$. Nevertheless, there exist some more solutions because of the
nonlinearity and the singularities of the differential equation
(\ref{g2y}).

\acknowledgments
We would like to thank Peter Baekler, Friedrich W.~Hehl, Alfredo
Mac{\' i}as (UAM, Mexico, D.F.), and Yuri N.~Obukhov for
useful comments. Moreover, the careful evaluation of E.~J.~Copeland
helped us to improve the content of our paper.
Research support for F.E.S.~was provided by the Deutsche
Forschungsgemeinschaft, project He $528/14-1$, whereas
E.W.M.~gratefully acknowledge support from Conacyt.

\begin{figure}
\caption[]{Case: $n_s=1.01$ up to $n_s=2.1$. The inflaton 
potential $U(\phi)$ is presented in units of $[1/(\kappa A)]$ and
the scalar field $\phi $ in units of $[\sqrt{2/\kappa }]$.}
\label{fig.0}
\end{figure}

\begin{figure}
\caption[]{Case: $n_s=1.1$.
It has the feature of the old inflationary
theory. One would get this potential if one find the same value $n_s$
for each wavelength $\lambda $.
The physical units are the same as in Fig.~\ref{fig.0}.}
\label{fig.1}
\end{figure}

\begin{figure}
\caption[]{Case: $n_s=10$.
It describes a potential of new inflationary theory. Again, this
potential is valid, if $n_s$ is independent of the wavelength
$\lambda $. The physical units are the same as in Fig.~\ref{fig.0}.}
\label{fig.2}
\end{figure}

\begin{figure}
\caption[]{Case: $0.01<n_s<0.95$ and $\lambda /\lambda_0=n_s^2$.
The physical units are the same as in Fig.~\ref{fig.0}.}
\label{fig.3}
\end{figure}

\begin{figure}
\caption[]{Case: $1.01<n_s<2.1$ and $\lambda /\lambda_0=1/n_s^2$.
The physical units are the same as in Fig.~\ref{fig.0}.}
\label{fig.4}
\end{figure}

\begin{figure}
\caption[]{Case: $1.01<n_s<2.1$ and $\lambda /\lambda_0=1/\sqrt{n_s}$.
The physical units are the same as in Fig.~\ref{fig.0}.}
\label{fig.5}
\end{figure}

\begin{figure}
\caption[]{Several potentials $U(\phi )$ in units of
$[1/(\kappa )]$ depending on the scalar field $\phi $ in units of
$[\sqrt{2/\kappa }]$. We have chosen the initial conditions:
a) $g(5)=-0.1$, $g'(5)=1$, and $\phi (5)=1$ for the drawn curve,
b) $g(5)=-0.1$, $g'(5)=0.1$, and $\phi (5)=1$ for the dashed curve, and
c) $g(5)=-1$, $g'(5)=1$, and $\phi (5)=1$ for the dotted curve.
The solutions were found within the $H$-interval $[1,5]$.
For $\phi >1$, the minus sign was used in (\ref{4gy}),
whereas for $\phi <1$, the plus sign has produced the graphs.
One recognizes no smooth transition at $\phi =1$.}
\label{fig.6}
\end{figure}

\input psfig

\psfig{figure=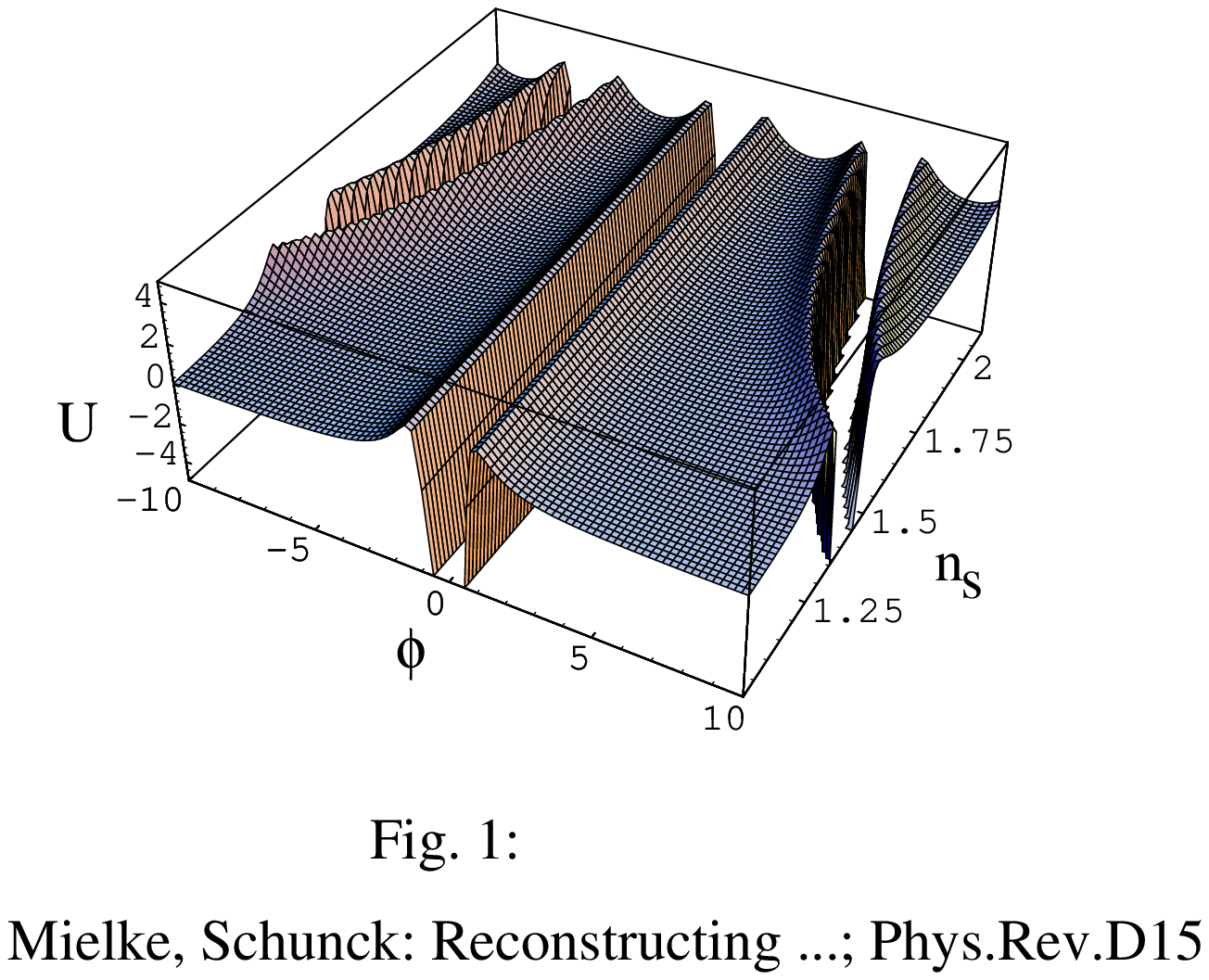}

\psfig{figure=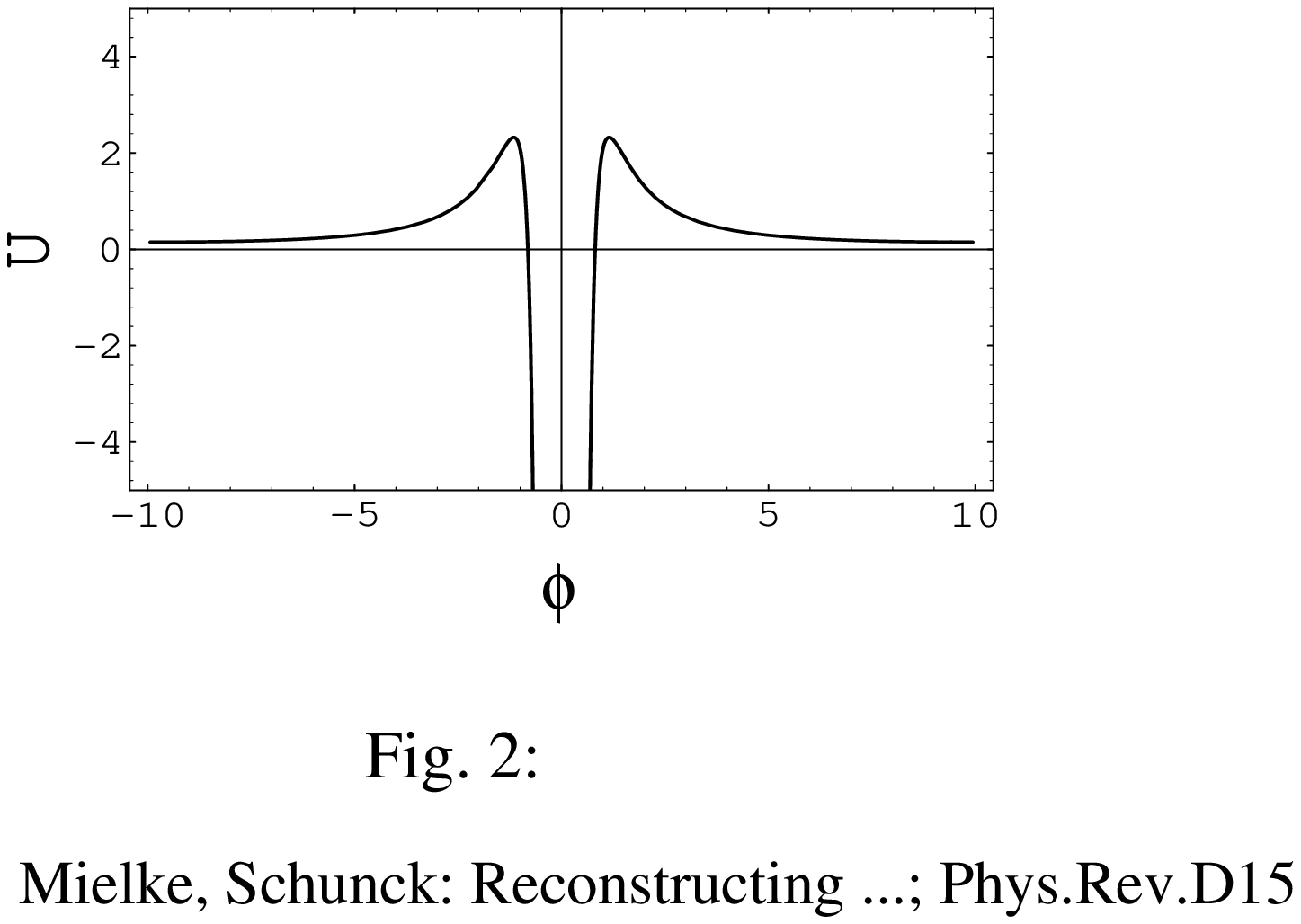}

\psfig{figure=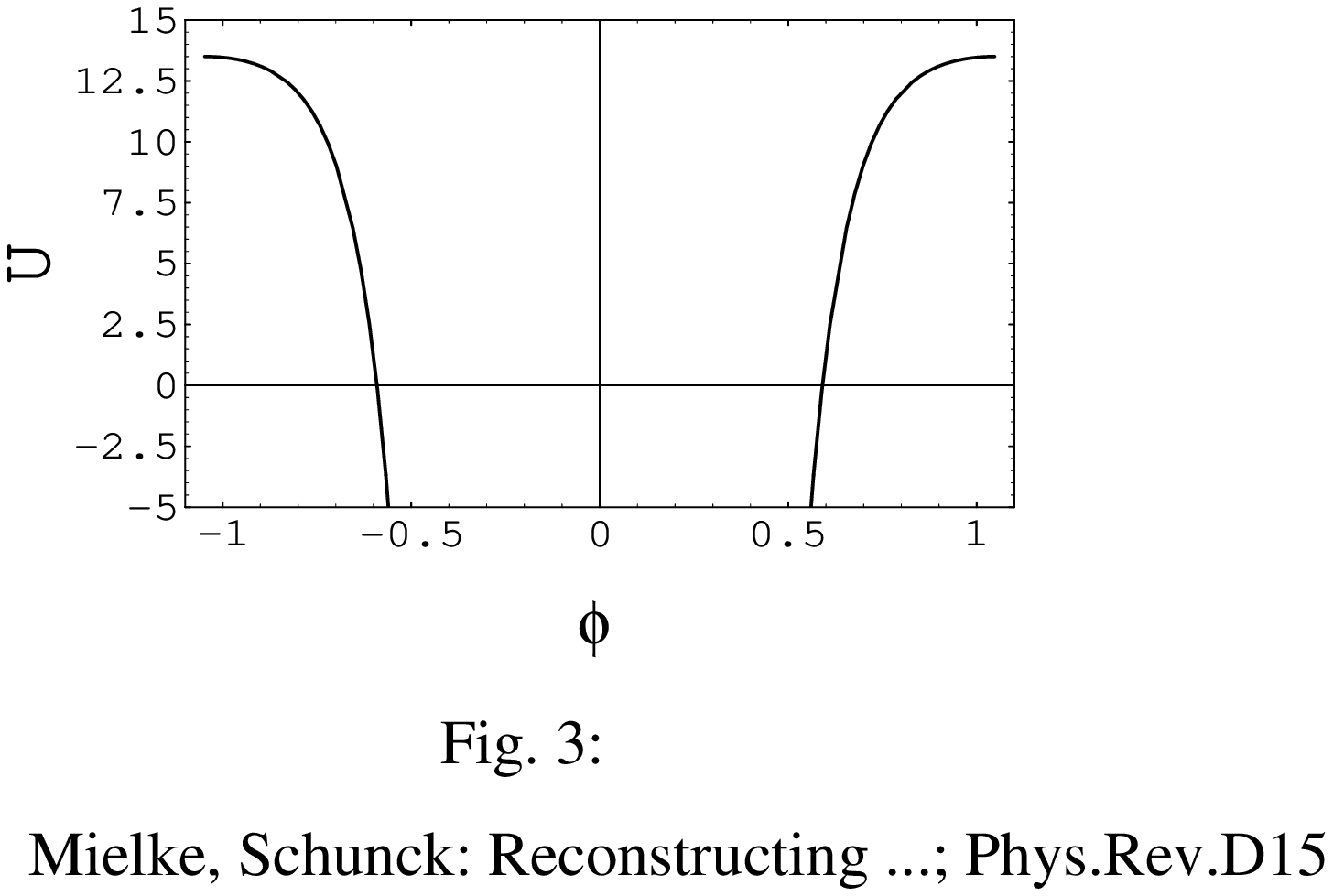}

\psfig{figure=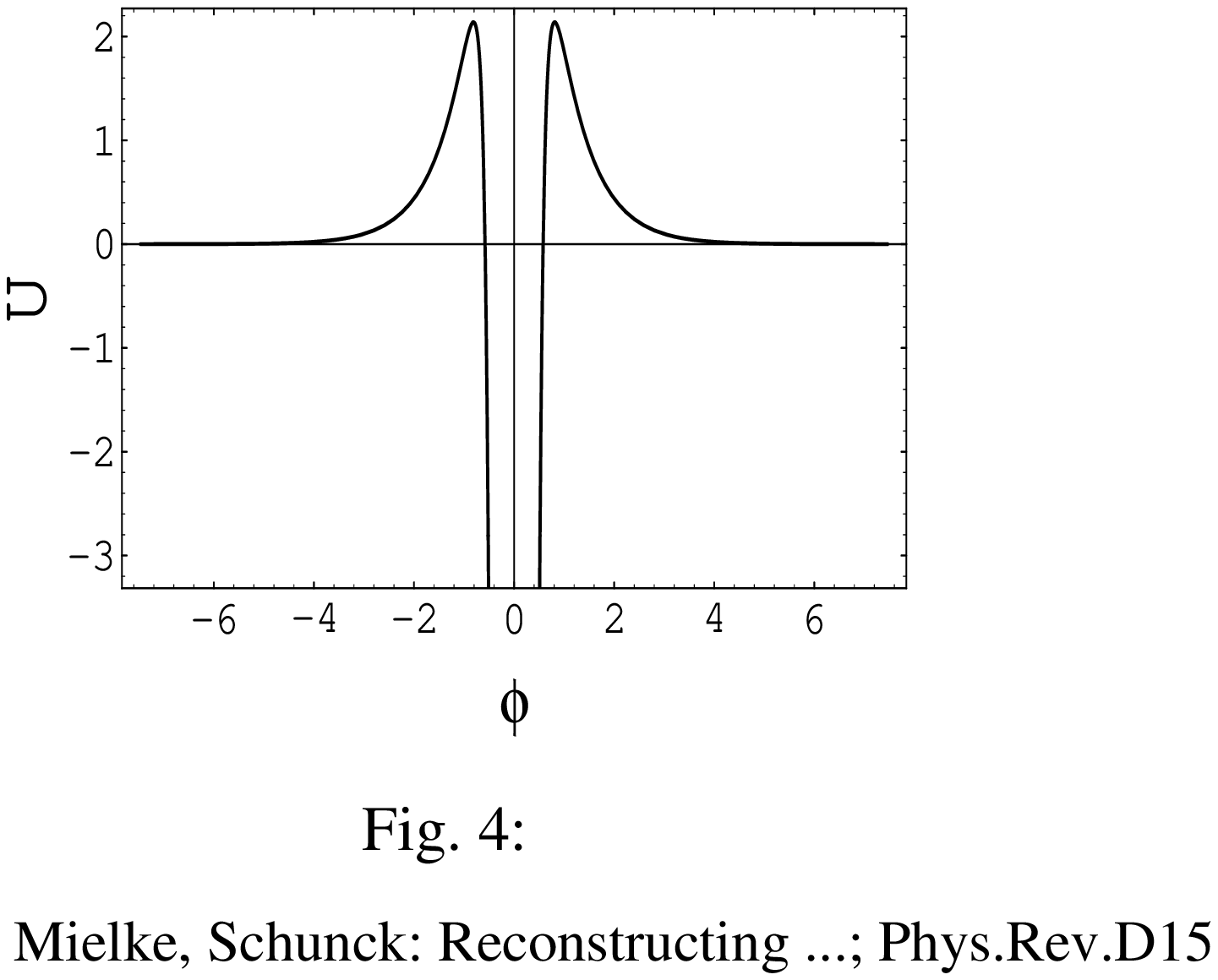}

\psfig{figure=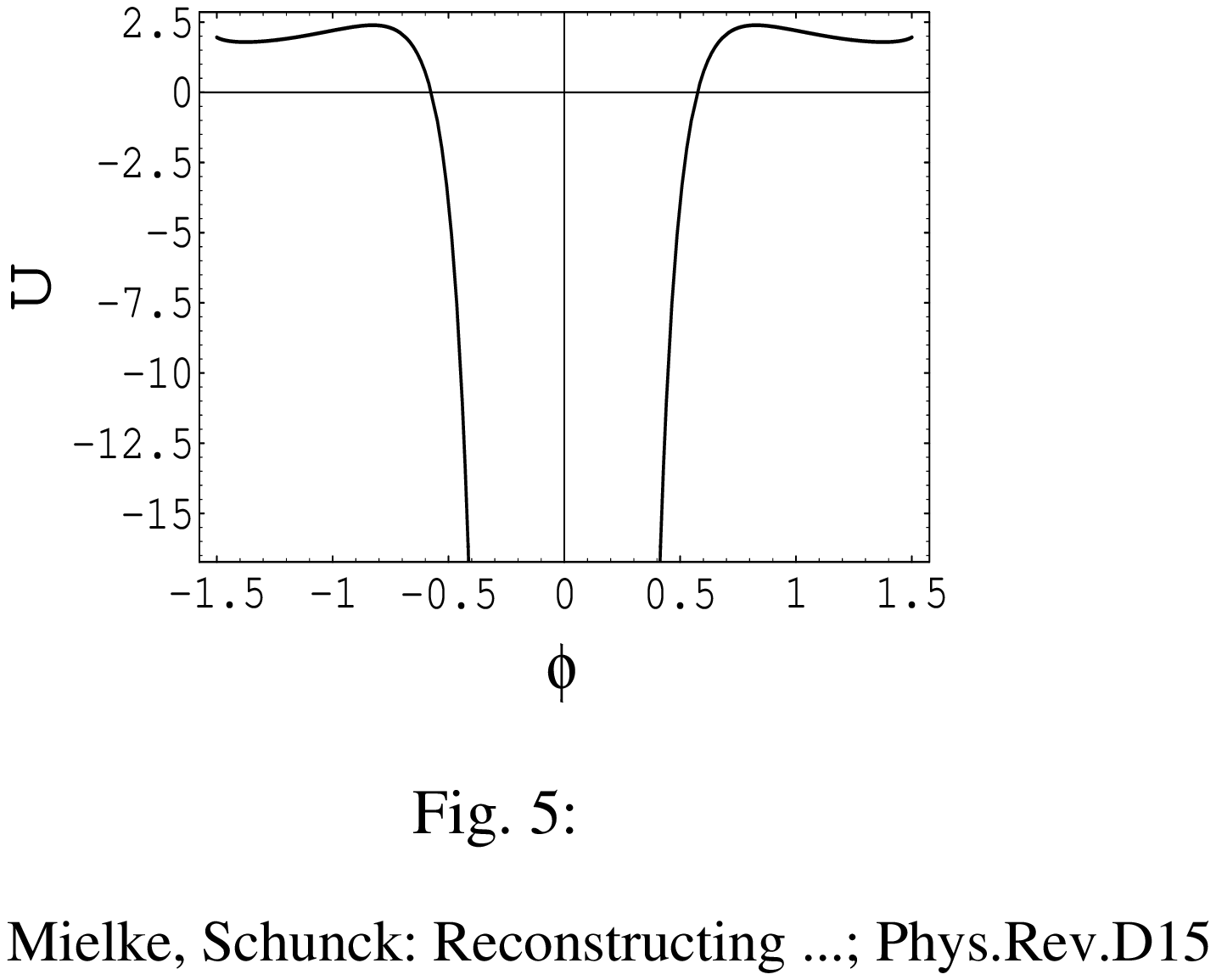}

\psfig{figure=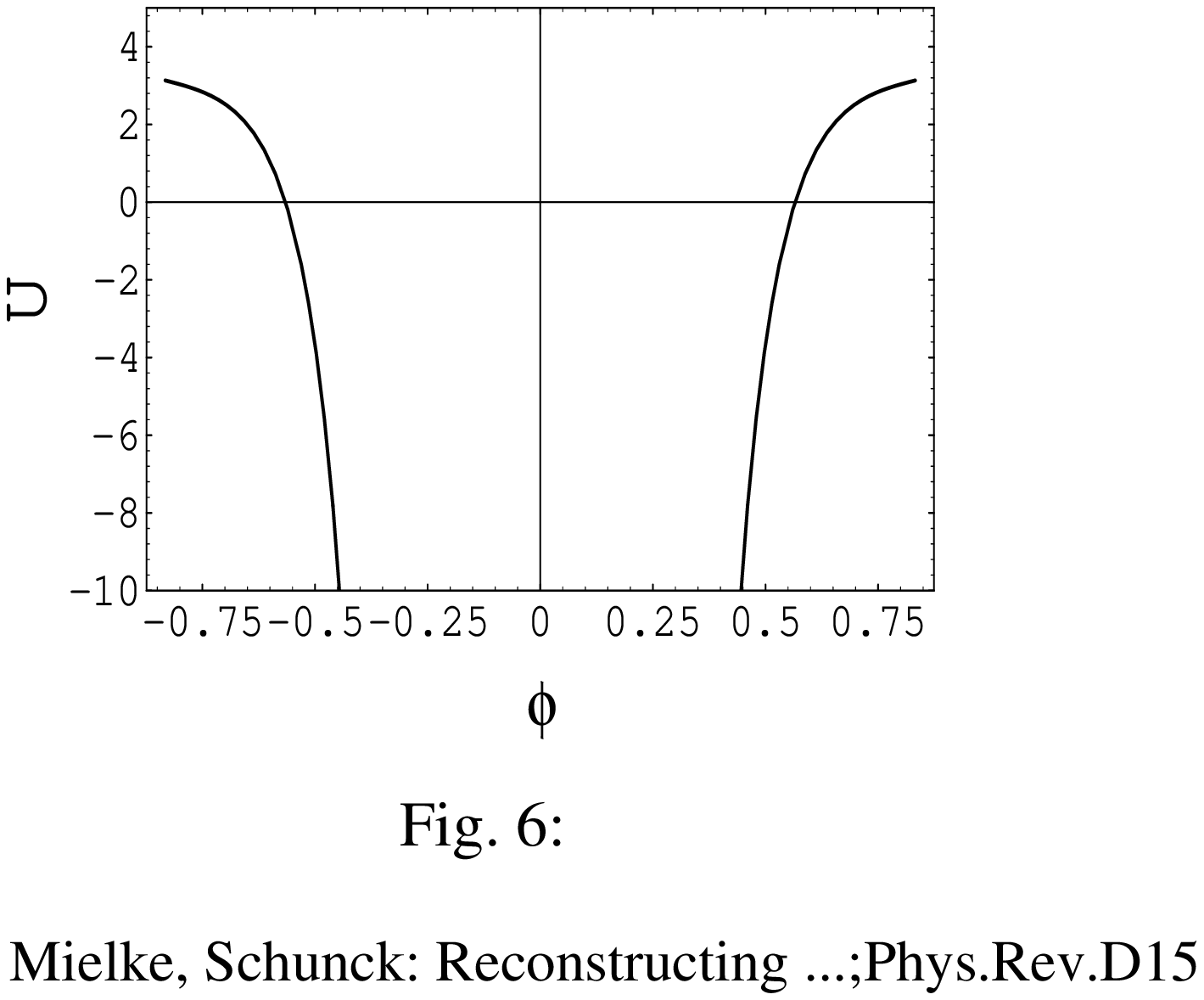}

\psfig{figure=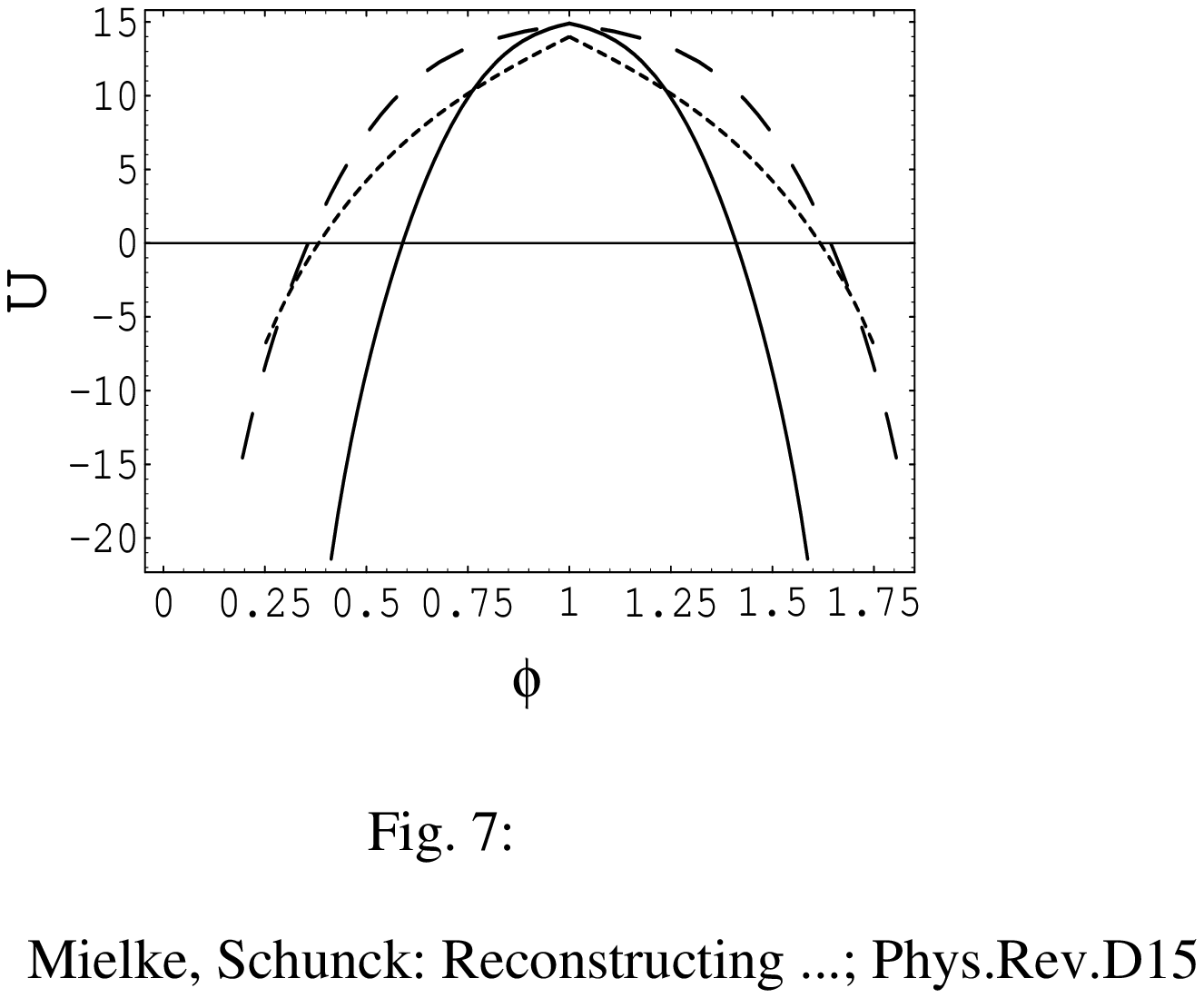}

\end{document}